\documentstyle[12pt]{article}
\begin{document}
\begin{titlepage}
\centerline{\bf Sparling two-forms, the conformal factor and}
\centerline{\bf the the gravitational energy density of the}
\centerline{\bf teleparallel equivalent of general relativity}
\bigskip
\centerline{\it J. W. Maluf$\,^{*}$}
\centerline{\it Departamento de F\'isica}
\centerline{\it Universidade de Bras\'ilia}
\centerline{\it C.P. 04385}
\centerline{\it 70.919-970  Bras\'ilia, DF}  
\centerline{\it Brazil}
\date{}
\begin{abstract}
It has been shown recently that within the framework of the
teleparallel equivalent of general relativity (TEGR) it is 
possible to define in an unequivocal way the energy density 
of the gravitational field. The TEGR amounts to an {\it alternative
formulation} of Einstein's general relativity, not to an alternative
gravity theory. The localizability of the gravitational energy has
been investigated in a number of spacetimes with distinct topologies, 
and the outcome of these analises agree with previously known
results regarding the exact expression of the gravitational energy,
and/or with the specific properties of the spacetime manifold.
In this article we establish a relationship between the expression
for the gravitational energy density of the TEGR and the Sparling
two-forms, which are known to be closely connected with the 
gravitational energy. We also show that our expression of 
energy yields the correct value of gravitational mass contained
in the conformal factor of the metric field.
\end{abstract}
\thispagestyle{empty}
\vfill
\par
\noindent (*) e-mail: wadih@guarany.cpd.unb.br
\end{titlepage}
\newpage

\noindent {\bf I. Introduction}\par
\bigskip

The problem of a consistent definition of the gravitational energy
has been addressed over the years by means of rather different 
methods. The traditional approach amounts to defining the total
gravitational energy of an asymptotically flat spacetime 
through the use of  pseudotensors. 
A closely related approach consists in associating 
the gravitational energy with the surface term that appears
in the total Hamiltonian of the gravitational field. An alternative
way of defining  the gravitational energy is provided by
the concept  of quasi-local energy. The quasi-local definition of
energy, momentum and angular momentum associates these quantities
with an arbitrary two-surface $S$ of an arbitrary spacetime
manifold. This approach is presented in ref.\cite{Brown}, where a
comprehensive list of references on the problem of energy
in general relativity is presented.

A recent approach to the problem of the definition of the gravitational
energy has arisen in the framework of the teleparallel equivalent of
general relativity (TEGR)\cite{MalufGRG,Maluf1,Maluf2}.
The latter is an alternative
formulation of general relativity. This formulation is established by
means of the tetrad field $e_{a\mu}$ and the spin connection 
$\omega_{\mu ab}$, which are totally independent field quantities, even 
at the level of field equations. The metric tensor obeys Einstein's
equations. However, the action integral is constructed in terms of
the torsion tensor. The Lagrangian density of the TEGR is a functional
of the  torsion tensor, but is precisely equivalent to the ordinary
scalar density $eR(e_{a\mu})$, provided the curvature tensor
$R_{ab \mu \nu}(\omega)$ vanishes. In fact the vanishing of
$R_{ab \mu \nu}(\omega)$ is one basic motivation for considering the TEGR,
since this property leads to the establishment of a {\it reference
space}\cite{Maluf3}.

The definition of the gravitational energy {\it density} emerges in the
context of the Hamiltonian formulation of the TEGR. The latter is
considered in ref.\cite{Maluf1}. It has been shown that, under a suitable
gauge fixing, the Hamiltonian formulation of the TEGR is well defined,
as the constraints turn out to be first class. The major property
of the Hamiltonian formulation is that the Hamiltonian constraint
$C=0$ can be written as 

$$C\;=\;H\,-\,E_{ADM}\;=\;0\;,$$

\noindent in the case of asymptotically flat 
spacetimes\cite{MalufGRG,Maluf2}.
$H$ is interpreted as the effective Hamiltonian. The 
Arnowitt-Deser-Misner (ADM) energy\cite{ADM} 
($E_{ADM}$) arises in the Hamiltonian constraint upon integration
over the whole three-dimensional spacelike hypersurface of a scalar
density $\varepsilon(x)$, which can be written as a total divergence
of the form ${1\over{8\pi G}}\partial_i(eT^i)$, 
where $T^i$ is the trace of the torsion
tensor. We propose that $\varepsilon(x)$ represents the gravitational
energy density for spacetimes with any topology. In fact we
have applied it to the determination of the distribution of 
gravitational energy in de Sitter space and found that the resulting
analysis is in total agreement with the phenomenological features
of this space\cite{Maluf4}.
Therefore we assert that the Hamiltonian constraint
equation can be written as $C\,=\,H-E\,=0$ for {\it any} spacetime.

The existence of the gravitational energy density as given above
allows the conclusion that the gravitational energy is localizable,
in the same way as the electromagnetic energy is also localizable.
The very idea of a black hole lends support to this conclusion. If
there exists an ammount of gravitational energy (mass) inside the
event horizon of a black hole, then by means of a coordinate 
transformation we do not expect to make this energy vanish.

In this article we establish a relationship between the 
gravitational energy density $\varepsilon(x)$ and the Sparling
two-forms. The latter are known to yield the gravitational energy
and momentum of an asymptotically 
flat spacetime by  means of an integration 
over a  two-surface at infinity. In fact the Sparling forms are
closely connected with various pseudotensors of energy-momentum 
of the gravitational field. We also show that under
integration $\varepsilon(x)$ captures the information about the
gravitational mass contained in the conformal factor of the metric 
field, provided the appropriate boundary conditions at infinity
are imposed. Thus our expression  of energy encompasses the 
features of distinct approaches to the definition of the 
gravitational energy.

We remark that although there exist in the literature several expressions 
for the definition of the {\it total} gravitational 
energy, the present expression is 
unique as it is the only one that possesses the feature of localizability.
It is precisely this feature that allows obtaining the striking 
result regarding rotating black holes (cf. ref.\cite{Maluf3}), namely, 
that the energy contained within the outer horizon of a Kerr black hole 
is essentially equal to  twice its irreducible mass.

Notation: spacetime indices $\mu, \nu, ...$ and local Lorentz indices
$a, b, ...$ run from 0 to 3. In the 3+1 decomposition latin indices 
from the middle of the alphabet indicate space indices according to
$\mu=0,i,\;\;a=(0),(i)$. The tetrad field $e^a\,_\mu$ and
the spin connection $\omega_{\mu ab}$ yield the usual definitions
of the torsion and curvature tensors:  $R^a\,_{b \mu \nu}=
\partial_\mu \omega_\nu\,^a\,_b +
\omega_\mu\,^a\,_c\,\omega_\nu\,^c\,_b\,-\,...$,
$T^a\,_{\mu \nu}=\partial_\mu e^a\,_\nu+
\omega_\mu\,^a\,_b\,e^b\,_\nu\,-\,...$. The flat spacetime metric 
is fixed by $\eta_{(0)(0)}=-1$. \\

\bigskip
\bigskip

\noindent {\bf II. The TEGR}\par
\bigskip
Throughout the paper we will consider only asymptotically flat
spacetimes, because it is in this geometrical framework that the 
Sparling two-forms are relevant for energy considerations.
The Lagrangian density of the TEGR in empty spacetime is given by 

$$L(e,\omega,\lambda)\;=\;-ke({1\over 4}T^{abc}T_{abc}\,+\,
{1\over 2}T^{abc}T_{bac}\,-\,T^aT_a)\;+\;
e\lambda^{ab\mu\nu}R_{ab\mu\nu}(\omega)\;.\eqno(1)$$

\noindent where $k={1\over {16\pi G}}$, $G$ is the gravitational 
constant; $e\,=\,det(e^a\,_\mu)$, $\lambda^{ab\mu\nu}$ are 
Lagrange multipliers and $T_a$ is the trace of the torsion tensor
defined by $T_a=T^b\,_{ba}$.   The tetrad field $e_{a\mu}$ and the
spin connection $\omega_{\mu ab}$ are completely independent field
variables.  The latter is enforced to satisfy the condition of
zero curvature. Therefore this Lagrangian formulation is in no way
similar to the usual Palatini formulation, in which the spin
connection is related to the tetrad field via field equations.

The equivalence of the TEGR with Einstein's general relativity is         
based on the identity

$$eR(e,\omega)\;=\;eR(e)\,+\,
e({1\over 4}T^{abc}T_{abc}\,+\,{1\over 2}T^{abc}T_{bac}\,-\,T^aT_a)\,-\,
2\partial_\mu(eT^{\mu})\;,\eqno(2)$$

\noindent which is obtained by just substituting the arbitrary
spin connection $\omega_{\mu ab}\, \equiv \,^o\omega_{\mu ab}(e)\,+\,
K_{\mu ab}$ in the scalar curvature tensor $R(e,\omega)$ in the
left hand side; $^o\omega_{\mu ab}(e)$ is the Levi-Civita 
connection and $K_{\mu ab}\,=\,
{1\over 2}e_a\,^\lambda e_b\,^\nu(T_{\lambda \mu \nu}+
T_{\nu \lambda \mu}-T_{\mu \nu \lambda})$ is the contorsion tensor.
The vanishing of $R^a\,_{b\mu\nu}(\omega)$, which is one of the
field equations derived from (1), implies the equivalence between
the scalar curvature $R(e)$, constructed out of $e^a\,_\mu$ only, 
and the quadratic combination of the torsion tensor. It also
ensures that the field equations arising from the variation of
$L$ with respect to $e^a\,_\mu$ are strictly equivalent to
Einstein's equations in tetrad form. Let 
${{\delta L}\over {\delta {e^{a\mu}}}}=0$ denote the field equations
satisfied by $e^{a\mu}$. It can be shown by explicit calculations
that

$${{\delta L}\over {\delta {e^{a\mu}}}}\;=\;
{1\over 2}\lbrace R_{a\mu}(e)\,
-\,{1\over 2}e_{a\mu}R(e)\rbrace\;.\eqno(3)$$

\noindent We refer the reader to
refs.\cite{Maluf1,Maluf2} for additional details.

In the Lagrangian density (1) we have not included the total divergence
$\partial_\mu(eT^\mu)$.  The reason is that the variation of the
latter in the action integral causes the appearance, via integration
by parts, of surface terms that do not vanish when $r \rightarrow \infty$.
In this limit we should
consider variations in $g_{\mu \nu}$ or in $e_{a\mu}$ that preserve
the asymptotic structure of the flat spacetime metric; the allowed
coordinate transformations must be of the Poincar\'e type. 
The variation of $\partial_\mu (eT^\mu)$ at infinity
under such variations of $e_{a\mu}$  does not vanish. 
Therefore the total divergence has to be dropped down.
On the other hand, all surface terms arising from 
partial integration in the variation of the action integral
constructed out of (1) vanish in the limit $r \rightarrow \infty$.
The action based solely on the quadratic torsion terms
does not require additional surface 
terms, as it is invariant under transformations that preserve the
asymptotic structure of the field quantities. This property fixes
the action integral, together with the requirement that the variation
of the latter must yield Einstein's equations, which is actually the 
case in view of (3) (the Hilbert-Einstein 
Lagrangian requires the addition of a surface term for the variation
of the action to be well defined; a clear discussion of 
this point is given in  ref.\cite{Faddeev}). 
Unfortunately in refs.[2-5] we have not given the correct argument for
the suppression of $\partial_\mu(eT^\mu)$. We argued, erroneously,
that the variation of this divergence vanishes at infinity
when we consider variations of $e_{a\mu}$ as defined above. This 
divergence has to be subtracted by hand, in the same way that 
the Hilbert-Einstein Lagrangian density requires an additional
surface term.

In what follows we will be interested in asymptoticaly flat spacetimes.
The Hamiltonian formulation of the TEGR can be successfully 
implemented if we fix the gauge $\omega_{0ab}=0$ from the 
outset, since in this case the constraints (to be 
shown below) constitute a {\it first class} set\cite{Maluf1}.
The condition $\omega_{0ab}=0$ is achieved by breaking the local
Lorentz symmetry of (1). We still make use of the residual time
independent gauge symmetry to fix the usual time gauge condition
$e_{(k)}\,^0\,=\,e_{(0)i}\,=\,0$. Because of $\omega_{0ab}=0$,
$H$ does not depend on $P^{kab}$, the momentum canonically 
conjugated to $\omega_{kab}$. Therefore arbitrary variations of
$L=p\dot q -H$ with respect to $P^{kab}$ yields 
$\dot \omega_{kab}=0$. Thus in view of $\omega_{0ab}=0$, 
$\omega_{kab}$ drops out from our considerations. The above 
gauge fixing can be understood as the fixation of a {\it global}
reference frame.

As a consequence of the above gauge fixing the canonical action 
integral obtained from (1) becomes\cite{Maluf2}

$$A_{TL}\;=\;\int d^4x\lbrace \Pi^{(j)k}\dot e_{(j)k}\,-\,H\rbrace\;,
\eqno(4)$$

$$H\;=\;NC\,+\,N^iC_i\,+\,\Sigma_{mn}\Pi^{mn}\,+\,
{1\over {8\pi G}}\partial_k (NeT^k)\,+\,
\partial_k (\Pi^{jk}N_j)\;.\eqno(5)$$

\noindent $N$ and $N^i$ are the lapse and shift functions, and 
$\Sigma_{mn}=-\Sigma_{nm}$ are Lagrange multipliers. The constraints
are defined by 

$$ C\;=\;\partial_j(2keT^j)\,-\,ke\Sigma^{kij}T_{kij}\,-\,
{1\over {4ke}}(\Pi^{ij}\Pi_{ji}-{1\over 2}\Pi^2)\;,\eqno(6)$$

$$C_k\;=\;-e_{(j)k}\partial_i\Pi^{(j)i}\,-\,
\Pi^{(j)i}T_{(j)ik}\;,\eqno(7)$$

\noindent with $e=det(e_{(j)k})$ and $T^i\,=\,g^{ik}e^{(j)l}T_{(j)lk}$. 
We remark that (4) and (5) are invariant under global SO(3) and
general coordinate transformations.  

We assume the asymptotic behaviour $e_{(j)k}\approx \eta_{jk}+
{1\over 2}h_{jk}({1\over r})$ for $r \rightarrow \infty$. In view
of the relation

$${1\over {8\pi G}}\int d^3x\partial_j(eT^j)\;=\;
{1\over {16\pi G}}\int_S dS_k(\partial_ih_{ik}-\partial_kh_{ii})
\; \equiv \; E_{ADM}\;\eqno(8)$$

\noindent where the surface integral is evaluated for 
$r \rightarrow \infty$, the integral form of 
the Hamiltonian constraint $C=0$ may be rewritten as

$$\int d^3x\biggl\{ ke\Sigma^{kij}T_{kij}+
{1\over {4ke}}(\Pi^{ij}\Pi_{ji}-{1\over 2}\Pi^2)\biggr\}
\;=\;E_{ADM}\;.\eqno(9)$$

\noindent The integration is over the whole three dimensional
space. Given that $\partial_j(eT^j)$ is a scalar  density,
from (7) and (8) we define the gravitational
energy density enclosed by a volume V of the space as

$$E_g\;=\;{1\over {8\pi G}}\int_V d^3x\partial_j(eT^j)\;.\eqno(10)$$  

\noindent It must be noted that $E_g$ depends only on the triads
$e_{(k)i}$ restricted to a three-dimensional spacelike hypersurface;
the inverse quantities $e^{(k)i}$ can be written in terms of 
$e_{(k)i}$. From the identity (3) we observe that the dynamics of
the triads does not depend on $\omega_{\mu ab}$. Therefore $E_g$
given above does not depend on the fixation of any gauge for
$\omega_{\mu ab}$.   

The reference space which defines the zero of energy has been 
discussed in ref.\cite{Maluf3}. Here we will 
briefly present the main ideas 
about its fixation. The establishment of a reference space 
requires the concepts of holonomic and anholonomic transformations
of coordinates. Let us consider the Euclidean space with metric
$\eta_{(i)(j)}=(+++)$, which is the spatial section of the 
Minkowski metric. We introduce coordinates $q^{(i)}$ such that the
line element of the Euclidean space is written as
$ds^2=\eta_{(i)(j)}dq^{(i)}dq^{(j)}$. We consider next a coordinate
transformations $dq^{(i)}=e^{(i)}\,_j(x)dx^j$, which allows us to 
rewrite $ds^2=\eta_{(i)(j)}e^{(i)}\,_m(x) e^{(j)}\,_n(x) dx^m dx^n=
g_{mn}dx^m dx^n$.  This transformation can be holonomic or 
anholonomic.

If the relation $dq^{(i)}=e^{(i)}\,_j(x)dx^j$ can be integrated
over the whole three-dimensional space, the transformation
$q^{(i)} \rightarrow x^j $ corresponds to a single-valued global
transformation, and therefore it is called holonomic. Both sets of
coordinates, $\lbrace q^{(i)} \rbrace$ and $ \lbrace x^j \rbrace$,
describe the three-dimensional Euclidean space, and we have
necessarily that $e^{(i)}\,_j$ is a gradient vector, i.e., 
$e^{(i)}\,_j={{\partial q^{(i)}}\over {\partial x^j}}$. However,
in general $dq^{(i)}=e^{(i)}\,_j dx^j$ cannot be globally integrated,
since $e^{(i)}\,_j$ may not be written as the gradient of a function,
namely, $e^{(i)}\,_j$ may not be of the type $e^{(i)}\,_j =
\partial_j q^{(i)}$. If the quantities $e^{(i)}\,_j $ are such that
$\partial_j e^{(i)}\,_k-\partial_k e^{(i)}\,_j \ne 0$, then the
transformation is called anholonomic.

For triads which are gradient vectors, the torsion tensor
$T_{(i)jk}=\partial_j e_{(i)k}-\partial_k e_{(i)j}$ vanishes identically.
A crucial result is that $T_{(i)jk}$ vanishes if and only if 
$\lbrace e^{(i)}\,_j \rbrace$ are gradient vectors\cite{Schouten}.
In the framework of the Hamiltonian formulation of the TEGR the
gravitational field corresponds to a configuration for which
$T_{(i)jk} \ne 0$. We conclude that {\it every} gravitational field
is {\it anholonomically} related to the three-dimensional 
Euclidean space, which is to be taken as the reference space.
Since the torsion tensor vanishes for the latter, as we have seen,
the gravitational energy of the reference space is zero.   \\

\bigskip
\bigskip

\noindent {\bf III. The Sparling two-forms and its relation with
the TEGR}\par
\bigskip

The Sparling two-forms $\sigma_a$ are defined by\cite{Sparling}
                                                
$$\sigma_a\;=\;-{1\over 2}\varepsilon_{abcd}\,\Gamma^{bc}
\wedge e^d\;,\eqno(11)$$

\noindent where $\varepsilon_{abcd}$ is the totally antisymmetric
Levi-Civita tensor such that $\varepsilon_{(0)(1)(2)(3)}=1\;;$ 
$\Gamma^{ab}$  and $e^a$ are one-forms, 
$\Gamma^{ab}= \Gamma_\mu \,^{ab}dx^\mu \;$,
$e^a=e^a\,_\mu dx^\mu $, which are related by

$$de^a\;+\;\Gamma^a\,_b\,\wedge \,e^b\;=\;0\,.\eqno(12)$$

\noindent In components, $\Gamma_{\mu ab}$ turns out to be the
Levi-Civita connection:

$$\Gamma_{\mu ab}\;=\;-{1\over 2} e^c\,_\mu(\Omega_{abc}\,-\,  
\Omega_{bac}\,-\,\Omega_{cab})$$

$$\Omega_{abc}\;=\;e_{a \lambda}(e_b\,^\nu \partial_\nu e_c\,^\lambda -
e_c\,^\nu \partial_\nu e_b\,^\lambda)\;.$$

\noindent It is known that in a coordinate basis which is 
asymptotically cartesian the Sparling forms are related to
various pseudotensors\cite{Wallner,Goldberg}. In particular,
it allows a definition of the {\it total} energy-momentum
$P_a$ of the gravitational field\cite{Goldberg}:

$$P_a\;=\;
-{1\over {8\pi G}} \oint _{\partial \Sigma} \sigma_a \;,\eqno(13)$$

\noindent where $\partial \Sigma$ actually represents a spacelike surface
$S$ at infinity ($\omega^{ab}$ of ref.\cite{Goldberg} differs by 
an overall sign from $\Gamma^{ab}$).         
We mention that the connection between the Sparling
forms with the Brown-York expression for quasi-local energy
has been investigated in ref.\cite{Lau}.

In order to establish the relation between $P_{(0)}$ given by (13) and
expression (10) we need to rewrite $\sigma_{(0)}$ in components. We
have:

\begin{eqnarray}
\sigma_{(0)} & =  & -{1\over 2}\varepsilon_{(0)bcd}\Gamma^{bc}
\wedge e^d \nonumber \\
& = & -{1\over 2}\varepsilon_{(i)(j)(k)}\Gamma_\mu\,^{(i)(j)}\,
e^{(k)}\,_\nu  dx^\mu \wedge dx^\nu \nonumber \\
& = & -{1\over 2}\varepsilon_{(i)(j)(k)}\Gamma_m\,^{(i)(j)} e^{(k)}\,_n
dx^m \wedge dx^n \;.\nonumber 
\end{eqnarray}

\noindent The last equality is obtained in view of the fact that 
$\sigma^{(0)}$ is integrated over a spacelike surface.

Let us introduce the surface element $dS_i$ defined by 

$$dS_i\;=\;{1\over 2}\varepsilon_{ijk} dx^j \wedge dx^k\;,$$

\noindent where $\varepsilon^{123}=1$. In view of the relations

$$e^{(i)}\,_m e^{(j)}\,_n e^{(k)}\,_l\, \varepsilon^{mnl}\;=\;
e\, \varepsilon^{(i)(j)(k)}\,$$

$$e^{(k)}\,_m \,\varepsilon^{mnl}\;=\;
e \, e_{(i)}\,^n e_{(j)}\,^l \varepsilon^{(k)(i)(j)}\;,$$

\noindent we find that $\sigma_{(0)}$ can be rewritten, after a number
of manipulations, as

$$\sigma_{(0)}\;=\;-e\,e^{(i)m} e^{(j)n}\,\Gamma_{n(i)(j)} 
dS_m\;.\eqno(14)$$

\noindent Therefore substitution of (14) in (13) leads to

$$P_{(0)} \; = \; {1\over {8\pi G}} \oint_S 
e^{(i)m} e^{(j)n}\Gamma_{n(i)(j)} dS_m \;=\;
{1\over {8\pi G}} \int_V \partial_m (e\,e^{(i)m} e^{(j)n}
\Gamma_{n(i)(j)}) d^3x \;. \eqno(15)$$

\noindent It is not difficult to verify that if we assume the 
asymptotic behaviour $e_{(i)j} \simeq \eta_{ij} + 
{1\over 2}h_{ij}({1\over r})$ when $r \rightarrow \infty$, and
impose the usual time gauge condition $e_{(i)}\,^0=e^{(0)}\,_j =0$
(the latter are {\it tetrad} components of the four-dimensional 
spacetime), then the expression above yields the ADM energy.    

The relation of (15) with the energy expression (10) can now be 
established in a straightforward way. The equivalence between the
two expressions rests on the identity

$$\partial_k(e\,e^{(i)k}\,e^{(j)n}\Gamma_{n(i)(j)})\; \equiv \;
\partial_k (e\,T^k)\;,\eqno(16)$$

\noindent which can be verified by just substituting $\Gamma_{m(i)(j)}$
on the left hand side of the equation above. Note that because of the
time gauge condition, which is also imposed in the Hamiltonian 
formulation of the TEGR, $\Gamma_{m(i)(j)}$ is constructed out of the
triads components restricted to the three-dimensional spacelike
hypersurface.

The equivalence between (10) and (15) can, however, be established
only if the integration is performed over the whole three-dimensional
space. The reason is that $\sigma_a$ defined by (11) is 
normally considered a non-invariant quantity, as it transforms
inhomogeneously under $e^a\,_\mu(x) \rightarrow 
\tilde e^a\,_\mu(x)=\Lambda^a\,_b(x)\,e^b\,_\mu(x)$, where 
$\Lambda^a\,_b(x)$ belongs to the $local\;\;SO(3,1)$.
On the other hand, in the framework of the TEGR 
$\partial_i(e\,T^i)$ is a scalar density, invariant under $ global\;
\;SO(3)$ transformations. This is a necessary requirement in order to
arrive at a Hamiltonian formulation with only first class constraints. \\

\bigskip
\bigskip
\noindent {\bf IV. The conformal factor of the metric}\par

\bigskip
An alternative way of defining the energy of an asymptotically flat 
gravitational field is by identifying it with the $O(r^{-1})$ part
of the conformal factor of the metric\cite{Brill}. However, this
identification is only possible if some boundary contitions are imposed.
Suppose that the metric field satisfies the following conditions 
in the asymptotic limit $r \rightarrow \infty$:

$$g_{ij}=\psi^4\,\delta_{ij}+\tilde h_{ij}\;;\; \psi=1+O(r^{-1})\;,$$

$$\tilde h_{ij}=O(r^{-1})\;;\;tr_{\delta}\, \tilde h_{ij}=O(r^{-2})\;;\;
\partial_j \tilde h_{ij}=O(r^{-3})\;.\eqno(17)$$

\noindent If this asymptotic behaviour is verified,
the following definition
is suggested for the energy of the gravitational field\cite{Brill}:

$$E\;=\;-{1\over {2\pi G}} \oint_{S \rightarrow \infty} 
dS_i\,\partial_i \psi\;.\eqno(18)$$

We will show that a similar statement can be made in the context of
expression (10). Let us consider the conformal factor $\psi$ 
as above, with the asymptotic behaviour $\psi =1+O(r^{-1})$, and write

$$e_{(k)i}\;=\;\psi^2\;^oe_{(k)i}\,+\,\tilde e_{(k)i}\;,\eqno(19)$$

\noindent where $\lbrace ^oe_{(k)i} \rbrace$ are triads of the flat
three-dimensional space (in cartesian coordinates, 
$^oe_{(k)i}=\delta_{ki}$) and $\tilde e_{(k)i}$ is such that for
$r \rightarrow \infty$ we have $\tilde e_{(k)i}=O(r^{-1})$. In terms of 
these quantities we can construct the trace of the torsion tensor
$T^i=g^{ik}\,e^{(l)j}\,T_{(l)jk}$:

$$T^i\;=\;g^{ik}\,e^{(l)j} \lbrack ^oe_{(l)k}\;\partial_j \psi^2
- ^oe_{(l)j}\; \partial_k \psi^2 + \partial_j {\tilde e_{(l)k}}-
\partial_k{\tilde e_{(l)j}} \rbrack \;.$$

In order to evaluate (10) and compare it with expression (18) we will
have to obtain the value of $eT^i$ on a surface $S$ at infinity. 
In this limit, the last two terms of $T^i$ of the expression above 
contribute  to the surface integral as

$$e^{(l)j}\,\partial_j {\tilde e_{(l)k}} \rightarrow 
\partial_j(\;^oe^{(l)j}\,{\tilde e_{(l)k}})\;,\eqno(20)$$

$$e^{(l)j}\,\partial_k {\tilde e_{(l)j}} \rightarrow
\partial_k (\;^oe^{(l)j} {\tilde e_{(l)j}})\;.\eqno(21)$$

\noindent We would like to have these quantities vanishing under
integration. For this purpose it is necessary to have them
falling off as $1\over r^3$ at infinity. This will be the case if
we require

$$\partial_j (\,^oe^{(l)j}\,e_{(l)k})=O(r^{-3})\;;\;^oe^{(l)j}
\tilde e_{(l)j}=O(r^{-2})\;.\eqno(22)$$

\noindent We observe that conditions (22) are the equivalent in
triad form of conditions (17): the first condition above is equivalent
to $\partial_j \tilde h_{ij}=O(r^{-3})$, the second one to
$tr_{\delta}\,\tilde h_{ij}=O(r^{-2})$.
Therefore we assume them to hold,
together with (19), for asymptotically flat gravitational fields.
As a consequence we obtain 

$${1\over {8\pi G}}\oint_{S \rightarrow \infty} dS_i \;eT^i\;=\;
-{1\over {2\pi G}}\oint_{S\rightarrow \infty}dS_i\partial_i\,\psi\;,
\eqno(23)$$

\noindent which establishes the equivalence with (18).    

As a simple application of expression (23) let us consider
the Schwarzschild metric in isotropic coordinates. The metric
for the spacelike hypersurface is given by

$$ds^2\;=\;\biggl( 1+{m\over {2r}}\biggr)^4(dr^2 
+r^2\,d\theta^2 +r^2\,sin^2\theta\,d\phi^2)\;,\eqno(24)$$

\noindent where, for convenience, we make use of spherical coordinates.
We have also made $G=1$. The triads associated with this metric read

$$e_{(k)i}\;=\;\pmatrix{\psi^2\,sin\theta cos\phi &
r\,\psi^2\,cos\theta cos\phi & -r\,\psi^2\,sin\theta sin\phi\cr
\psi^2\,sin\theta sin\phi & r\,\psi^2\,cos\theta sin\phi &
r\,\psi^2\,sin\theta cos\phi\cr
\psi^2\,cos\theta& -r\,\psi^2\,sin\theta & 0\cr}\eqno(25)$$

\noindent where $\psi=(1+{m\over {2r}})$.
In the expression above $(k)$ and $i$ are line and
column indexes, respectively. It is easy to verify that 
$e^{(k)}\,_i e_{(k)j}$ yields precisely the metric components of
the line element (24).
We will exempt from presenting the details of the calculations, 
which in fact are not complicated. The only contribution to the
surface integral is given by

$$eT^1\;=\;eg^{1j}\,e^{(k)i}\,T_{(k)ij}\;=\;
-4\,r^2\,sin\theta\,\psi {{\partial \psi}\over {\partial r}}\;.$$

\noindent We find (recall that $G=1$)

$$E\;=\;{1\over {8\pi}}\oint_{S\rightarrow \infty} dS\,eT^1\;=\;
{1\over {8\pi}}\int d\theta\,d\phi\,sin\theta\,2m\;=\;m\;,\eqno(26)$$

\noindent as expected.  \\

\noindent {\bf V. Comments}\par
\bigskip
We have seen that expression (10) encompasses  the features of previous,
distinct approaches to the definition of the gravitational energy. 
Furthermore,
it is still possible to consider {\it localized} gravitational energy.
In this paper we have considered the definitions of gravitational
energy  (i) constructed out of the Sparling forms, and (ii)
constructed out of the conformal factor of the metric field. 
None of these approaches arise in a natural or conventional way
in the Lagrangian or Hamiltonian formulations of general relativity.
Moreover, they provide only the {\it total} gravitational energy
of asymptotically flat  fields. We have seen that
these definitions are in fact related to the definition
(10) for the gravitational energy, which in turn does appear as one
element of the Hamiltonian constraint $C$ in the 
canonical formulation of the TEGR. Thus expressions (13) and (18)
can be understood as different manifestations of the energy
expression (10), when the integration is performed over the whole
three-dimensional space. This result supports the general validity of
expression (10).

\bigskip
\bigskip
\noindent {\it Acknowledgements}\par
\noindent This work was supported in part by CNPQ.

\end{document}